# Magnetic Transition in LaVO$_3$/LaTiO$_3$ superlattice: A DFT+MC study

Mukesh Sharma and Tulika Maitra

Department of Physics, Indian Institute of Technology Roorkee, Roorkee, 247667, Uttarakhand, India

*Corresponding author: msharma1@ph.iitr.ac.in

**Abstract**

Magnetic phase transitions have been explored in a superlattice formed by stacking monolayers of LaTiO$_3$ and LaVO$_3$ alternately, using *ab-initio* density functional theory (DFT) and Monte-Carlo (MC) simulations. DFT derived intra-layer and inter-layer exchange interaction parameters were used for the MC simulations on a Ising spin model Hamiltonian. Two sharp peaks observed in specific heat without the interlayer exchange coupling indicate two independent magnetic ordering in LaTiO$_3$ and LaVO$_3$ layers at different temperatures. Inclusion of interlayer coupling leads to one sharp peak at higher temperature with a broad hump like feature at lower temperature in specific heat indicating a single magnetic phase transition *to C-type* antiferromagnetic phase in the superlattice.

## Introduction

Magnetic multilayers, where layers of different magnetic materials are stacked along a particular direction, are a part of a very active area of research because of their huge potential for applications in magnetic switch, memory devices, spintronics etc. Another important area of recent interest is oxide heterostructures where two different oxide perovskite materials (few layers of each) are grown on top of each other to form a heterointerface. These interfaces often show various exotic phenomena like superconductivity, metal-insulator phase transition, magneto resistance, etc[1], which are not present in their bulk counterparts. Such engineered structures formed by combining two different materials of varying properties to generate novel phenomena, is one of the major goals for the research community.

In this work we have studied the magnetic phase transition in a magnetic multilayer formed by two different transition metal oxide perovskite materials LaTiO$_3$ (LTO) and LaVO$_3$ (LVO). Oxide perovskites have general chemical formula ABO$_3$, where A and B are cations and O is the anion. The structure is shown in Fig. 1(a) where one can see that B$^{3+}$ ion is surrounded by 6 O$^{2-}$ ions in an octahedral geometry. In Fig. 1(b) we present the unit cell of LTO/LVO heterostructure we have considered here where we have stacked monolayers of LTO and LVO alternately along c-direction. With periodic boundary condition along all three directions we construct the multilayer LTO/LVO. Since both Ti$^{3+}$ and V$^{3+}$ are magnetic with spin S=1/2 and S=1, this system is a magnetic multilayer.

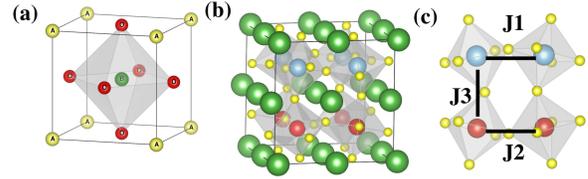

FIG. 1. (a) Perovskite ABO$_3$ structure where yellow, green, and red spheres represent A, B, and O ions respectively, (b) Unit cell of heterostructure LaTiO3/LaVO3 (c) J$_1$, J$_2$, J$_3$ showing the exchange coupling between Ti-Ti, V-V, Ti-V ions. In (b-c) green, blue, red, and yellow spheres represent La, Ti, V, and O ions respectively.

## Method

To explore the magnetic phase transition in our system, we have first estimated various magnetic exchange coupling (J) using Density Functional Theory (DFT) calculations. Vienna Ab-initio Simulation Package (VASP)[2] has been used for DFT calculations within the Local Density approximation (LDA) with finite Hubbard repulsion terms (on-site Coulomb(U), exchange energy(J)) using Dudarev[3] approach with U$_{eff}$ =(U-J) of values 8eV, 3eV and 3eV applied on *La(f), Ti(d)* and *V(d)* orbitals. We estimated three J values (J$_1$, J$_2$ and J$_3$) corresponding to the Ti-Ti, V-V and Ti-V interactions as shown in Fig. 1(c) as per the method suggested by H. J. Xiang et al[4].

To investigate the finite temperature magnetic phase transitions in our system we performed Monte-Carlo simulations on the following Ising spin Hamiltonian

$$H = \sum_{<ij>} J_1\, S^i_{Ti} S^j_{Ti} + J_2\, S^i_V S^j_V + J_3\, S^i_{Ti} S^j_V \quad (1)$$

where the sum runs over the nearest neighbor sites $i$ and $j$, and $J_1$, $J_2$ and $J_3$ are the exchange coupling energies between Ti-Ti, V-V, Ti-V ion pairs, and **S** is Ising spin with value 0.5(1.0) for Ti(V) ion.

For the MC, metropolis algorithm has been used. We used cubic lattice of size LxLxL (varying $8 \leq L \leq 30$ to reduce the size effects[5]) with periodicity along all directions. Out of $10^6$ Monte Carlo steps per spin (MCS), $5 \times 10^5$ MCS has been considered for thermal equilibration and next $5 \times 10^5$ MCS are used for calculating the observables like E(energy/sites), specific heat, etc. The specific is calculated by the formula $C_V = L^3 T^{-2}(<E^2> - <E>^2)$.

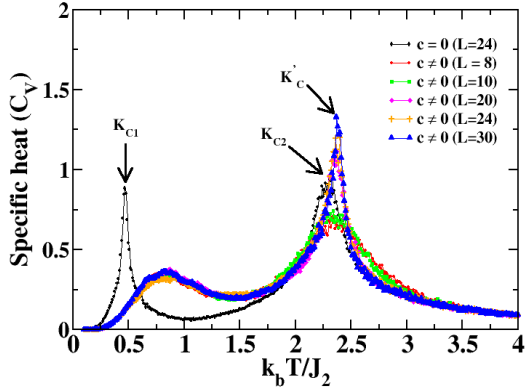

FIG. 2. Specific heat curves as a function of $k_b T/J_2$ shown here for lattice size (LxLxL).

## Results and Discussion

Using the lattice parameters from Eylem et al[6] who had studied the metal-insulator transitions in doped $LaTi_{1-x}V_xO_3$ bulk system, we have calculated $J_1$, $J_2$ and $J_3$ using DFT total energy calculations. Calculated $J_1$ and $J_2$ are observed to be antiferromagnetic (AFM) in nature with strengths **+33.07** and **+40.45** meV respectively whereas $J_3$ is found to be ferromagnetic (FM) in nature having strength **-11.83** meV and suggests a ***C-type*** AFM ordering at interface region. We have used these J values estimated from DFT in eq (1) to perform the MC simulations to explore the finite temperature properties. In the MC simulations we have normalized all the energy scales by $J_2$ which is the largest exchange coupling obtained and define $c=J_3/J_2$. We present in Fig 2 specific heat ($C_V$) as a function of normalized temperature ($k_b T/J_2$) for various lattice sizes without (c=0) and with interlayer coupling c. When there is no interlayer coupling (c=0) we observe two sharp peaks in the specific heat curve at $K_{c1}$ and $K_{c2}$ indicating two independent magnetic ordering in Ti and V layers at two different temperatures with V layers ordering at higher temperature $K_{c2}$(~2.33) and Ti layers are ordering at lower temperature $K_{c1}$(~0.47).

However, at finite interlayer exchange coupling (c≠0.0), we observe that the specific heat has one broad peak at lower temperature range which remains unchanged with respect to $L$ whereas the higher temperature peak is pushed to slightly higher temperature **K'$_c$** (~2.37) and gets sharper as we increase $L$. This sharp peak at higher temperature represents the critical temperature where the phase transition takes place in the LTO/LVO multilayer system whereas the broad peak indicating short range correlations. Our findings are consistent with previous results in other bilayer systems reported by L. Veiller et al[7].

## Conclusion

In summary, we have studied LTO/LVO magnetic multilayer system using a combination of DFT and MC calculations. We have calculated the intra-layer Ti-Ti and V-V as well as inter-layer Ti-V exchange interactions from DFT calculations. Using those exchange parameters, we have studied the magnetic phase transition in the multilayer system using MC simulations on a Ising type spin Hamiltonian. We observe that with finite interlayer exchange coupling two independent magnetic phase transitions in Ti and V layers are converge into a single phase transition for the entire system.

## Acknowledgement

MS acknowledges MHRD, India for research fellowship.